\documentclass[cpp,fleqn]{w-art}
\usepackage{times}
\usepackage{w-thm}
\theoremstyle{plain}

\theoremstyle{definition}

\usepackage[]{graphicx}
\chardef\bslash=`\\ 

\newcommand{\be}{\begin{equation}}
\newcommand{\ee}{\end{equation}}

\begin{document}

\DOIsuffix{theDOIsuffix}
\Volume{}
\Issue{}
\Month{}
\Year{}
\pagespan{1}{}
\Receiveddate{}
\Reviseddate{}
\Accepteddate{}
\Dateposted{}



\title{Bound States in Coulomb Systems \\
- Old Problems and New Solutions}

\author[Ebeling]{W. Ebeling$^1$\footnote{Corresponding
     author: e-mail: {\sf ebeling@physik.hu-berlin.de}}}
\address[\inst{1}]{Institut f\"ur Physik, Humboldt-Universit\"at zu Berlin,
Newtonstr. 15, D-12489 Berlin}
\author[Kraeft]{W.D. Kraeft$^{2,3}$\footnote{e-mail: {\sf
wolf-dietrich.kraeft@uni-rostock.de}}}
     \address[\inst{2}]{Institut f\"ur Physik, Ernst-Moritz-Arndt-Universit\"at Greifswald,
Felix-Hausdorff-Str.6, D-17487 Greifswald}
\author[R\"opke]{G. R\"opke$^3$\footnote{e-mail: {\sf
gerd.roepke@uni-rostock.de}}}
     \address[\inst{3}]{Institut f\"ur Physik, Universit\"at Rostock,
Universit\"atsplatz 3, D-18055 Rostock}
\begin{abstract}
We analyze the quantum statistical treatment of bound states in Hydrogen considered as a system of
electrons and protons. Within this physical picture we calculate isotherms of pressure for Hydrogen
in a broad density region and compare to some results from the chemical
picture.
First we resume in detail the two transitions along isotherms :\\
(i) the formation of bound states occurring by increasing the density from low to moderate values,\\
(ii) the destruction of bound states in the high density region, modelled here by
Pauli-Fock effects.\\
Avoiding chemical models we will show, why bound states according to a discrete part of the spectra
occur only in a valley in the T-p plane. First we study virial expansions in the canonical ensemble and then
in the grand canonical ensemble.
We show that in fugacity representations the population of bound states saturates at higher density
and that a combination of both representations provides quickly converging equations of state.
In the case of degenerate systems we calculated first the density-dependent energy levels,
and find the pressure in Hartree-Fock-Wigner approximation showing the prominent role of
Pauli blocking and Fock effects in the selfenergy.
\end{abstract}

\date{2nd draft 8.8.2011}

\maketitle

\section{Introduction}

In 1911, exactly hundred years ago, Rutherford invented a new model of matter,
which is the basis for the science dealing with strongly coupled Coulomb systems \cite{Rutherford}.
His model was based on the interpretation of existing experiments on the scattering of electrons on matter. In May 1911, Rutherford came forth with a model for the structure of atoms which provided a first understanding why electron scattering can so deeply penetrate into the interior of atoms, so far unexpected experimental results. Rutherford explained his scattering results as the passage of a high speed electron through an atom having a
positive central charge  $+N e$, and surrounded by a compensating charge of $N$ electrons \cite{Rutherford}.
In Rutherford's model the atom is nearly empty, it is made up of a central charge and electrons and the
glue keeping the system together are the Coulomb forces. Soon after, in 1913, a first quantum-mechanical treatment of the model was given by Bohr and a statistical treatment by Herzfeld.
According to the Bohr-Herzfeld theory the energies and the corresponding partition function are
\begin{eqnarray}
 E_{s}^* = - \frac{\mu e^4}{2 \hbar^2 s^2 }~, \qquad \mu = \frac{m_e m_+}{m_e + m_+} ~, \qquad
 \sigma(T) = \sum_{s} s^2 \exp\left( -\beta E_s \right) = \sum_{s=1}^{s_{\rm max}} s^2 \exp\left(\frac{I}{s^2 k_B T}\right)~.
 \label{eq:PlasmenZustandssumme1}
\end{eqnarray}
We denote the main quantum number by $s$ in order not to be mixed
up with the density $n$. According to the Bohr model there are infinitely many levels close to the series
limit $s \rightarrow \infty$ the terms increase as $s^2$ and the sum diverges. This is the result
for hydrogen, but the problem remains for all elements. The problem of the divergency of the atomic partition function took a long time before the mathematical
background for a serious treatment of this problem was available.
However urgent need to solve the ionization problems in astrophysics and
plasma physics forced fast solutions. For the calculations the problem was simplified
and the partition function was estimated by the first term of the sum, or by two or three terms.
Ignoring the open problems with the partition function, Eggert and Saha attacked the urgent problems of
astrophysics, considering only the ground state contribution.

An idea developed later was to cut the sum at the minimal terms $|E_s| \simeq k_B T$.
This way of solving the divergence problem is due to the classical paper of Planck \cite{Planck} published in 1924. Developing Plancks approach Brillouin proposed in 1930 a more smooth
way of removing the divergence which led to the formula \cite{Brillouin}
\begin{eqnarray}
\sigma_{\rm BPL}(T) =  \sum_{s=1}^{\infty} s^2 [\exp(\beta I/s^2) - 1 - (\beta I / s^2)]\,.
\label{BPLZustandssumme}
\end{eqnarray}
Using semi-classical methods, Planck could show that the (infinite) interaction
terms below and above the series limit
compensate each other.
This procedure was justified by methods of quantum statistics only in 1960 by Vedenov and Larkin \cite{Vedenov,Larkin} and subsequently by the present authors in collaboration with D. Kremp by using quantum-statistical
methods
\cite{EbAnnPhys,KrKr68,KrEbKr69,EbKrKrCPP70,KrKrEbPhy71,RedBook,EKKK76,GreenBook,EbKrKrRo86,KSKBook}.
We will show here that the Brillouin-Planck-Larkin procedure provides the most natural
way to avoid the divergencies for non-degenerate systems. Further we study degenerate systems and show that a consequent treatment of the identity of electrons in the Hartree-Fock approximation including bound states is a key for the study of hydrogenic bound states. The present approach is based on the method of Greens functions as developed in \cite{RedBook,EKKK76,GreenBook,EbKrKrRo86,KSKBook} and recently surveyed in a book dedicated to Nevill Mott \cite{MottBook}. The present investigation is concentrated on the so-called physical picture which stays within the Rutherford picture and avoids the introduction of chemical species like atoms, molecules etc.
In some of the earlier work the transition to a chemical picture was made following the principle of equivalence that bound states are to be treated on the same footing as free particles \cite{EbKrKrRo86,EbPhysica74}. This approach was quite successful in the description of partial ionization
\cite{RedBook,EbRi85,Pade90,Saumon,befjnrr,EbHaJRR5}. On the other hand the chemical picture may lead to principal difficulties e.g. if at high densities the minimum of the free energy is not well defined \cite{EbHiKr02,EbHaSp03}. This is the reason why we stay here within the physical picture trying to sum up the important higher order terms, this way following the line developed in several recent works \cite{AlastueyBaCoMa08,Alastuey,BoTrEb11,EbDubna}.

\section{Density and fugacity expansions for nondegenerate systems including nonlinear ring and Saha-type contributions}

\begin{figure}[htb]
\begin{center}
\includegraphics[width=4cm,angle=0]{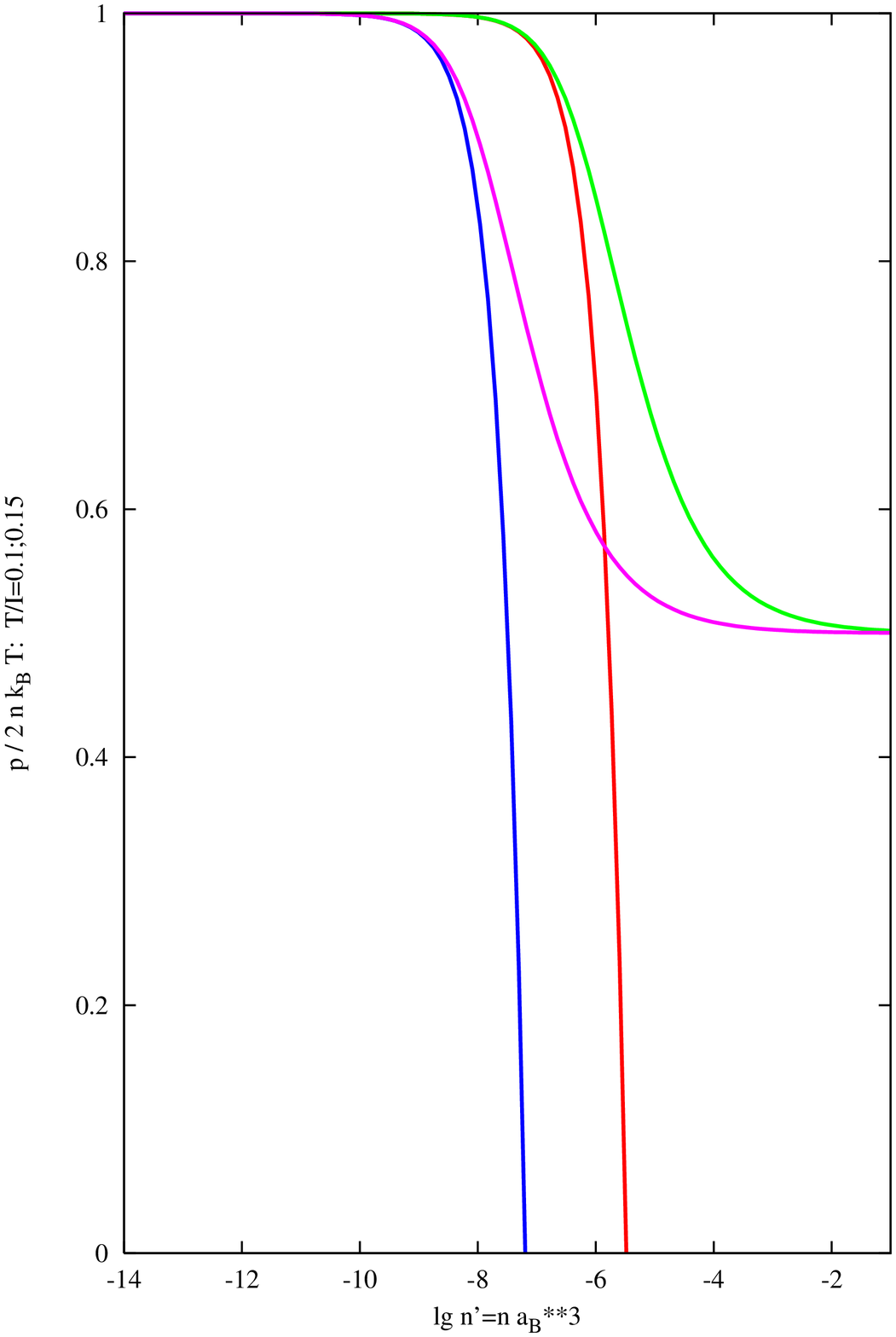}
\includegraphics[width=4cm,angle=0]{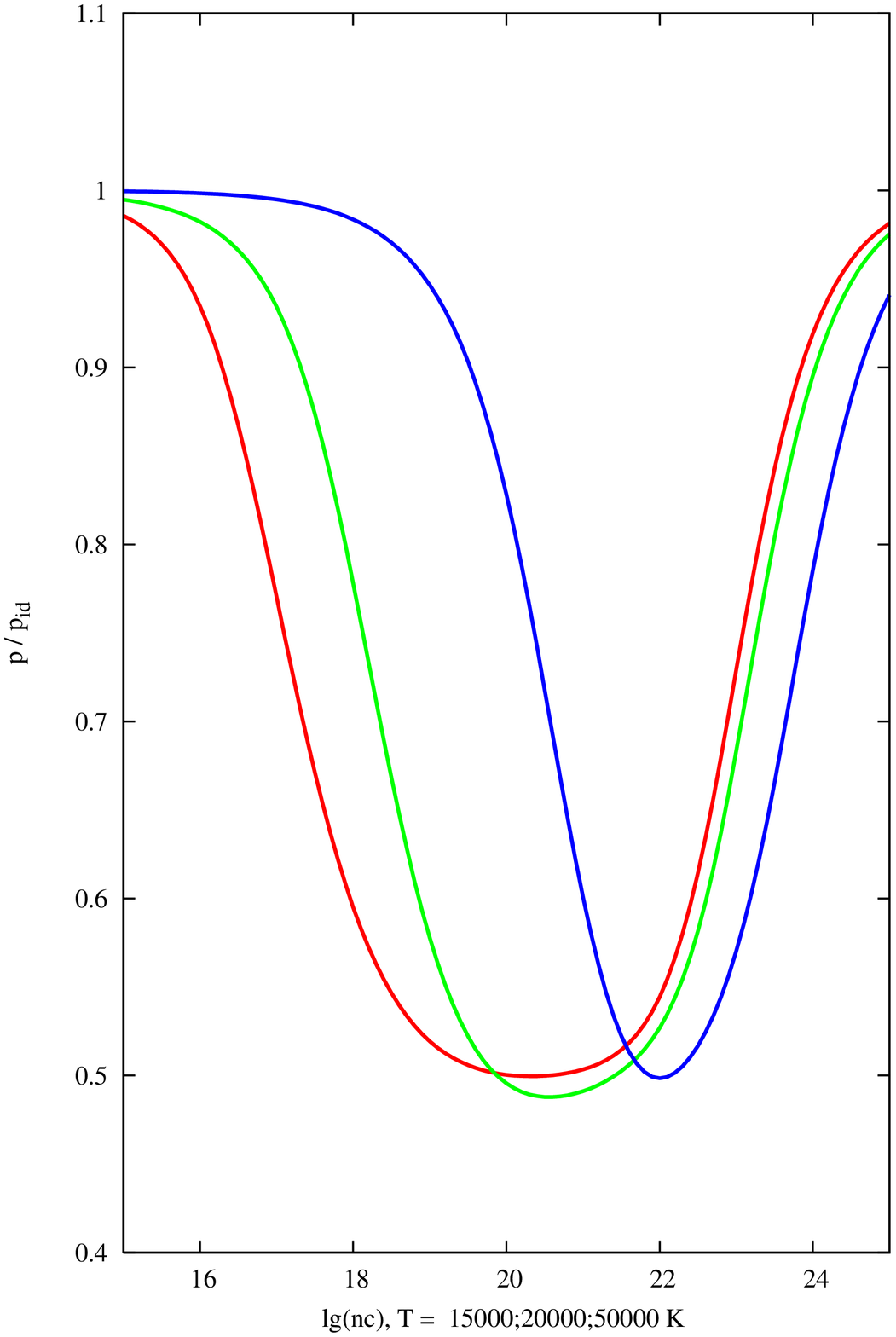}
\end{center}
\caption{Left panel: Pressure related to the total classical pressure $2 n k_B T$ at constant temperatures
$T = 15788 $ K and $T =23682$ K in dependence on the density (log-scale). The result of density expansions up to 2nd order virial terms (red and green) are compared to
corresponding fugacity expansions.
Right panel: Pressure related to the full ideal pressure (including the Fermi pressure) calculated from an extended density
representation including the ring sum $\phi(x)$ and the Saha function $A(x)$ for the temperatures $T = 15000$ K, 20000 K, 50000 K
in dependence on the density (in log-scale 15 - 25).}
\label{presse}
\end{figure}

Any successful description of plasmas has to go beyond the Saha theory which is something like the zeroth order approximation for plasmas. The derivation of Saha-type approximations from quantum statistics was first studied by Planck and Brillouin \cite{Planck,Brillouin} and is since then
a topic of permanent interest due to the importance of Saha-type equations for experimental work
and for many technological applications \cite{EbAnnPhys}.
We use here a systematic quantum statistical approach to the pressure on the basis of the Greens
function representations of the pressure \cite{RedBook,EKKK76,GreenBook,EbKrKrRo86}
\begin{eqnarray}
\label{qupress}
p(\beta, \mu_e,\mu_i)=p_{\rm id} - \frac{1}{2 V} \int_0^{1} \frac{d \lambda}{\lambda} \int d 1 {d {\tilde 1}}
V(1 {\tilde 1}) G_2 (1, {\tilde 1},1^{++},1^+: {\tilde t}_1 = t_1^+)
\end{eqnarray}
where $1 = \{\vec p_1, \sigma_1 \}$ denotes momentum and spin variables.
The Green's function representation works in the grand canonical ensemble which provides us a series
in the fugacities ($s_a$ - spin of particles, $a = e,i$)\\
$$
z_a = \frac{2s_a + 1}{\Lambda_a^3}\exp(\beta \mu_a) = n_a \exp(\beta
\mu_a^{\rm ex}); \qquad  \Lambda_a = \frac{h}{\sqrt{2 \pi m_a k_B T}}.$$
Note that $\mu_a^{\rm ex}$ is the excess part of the chemical potential (the part beyond the ideal Boltzmann term)
and that therefore $z_a \rightarrow n_a$ for small densities.
A representation by a density series may be obtain by inversion, i.e. by excluding the fugacities.
We summarize the existing results about the pressure of a hydrogenic quantum plasma as a function of the density
in a rather special form which
is asymtotically valid in the region of low and moderate temperatures $T \le I / k_B T$. Using for this region the
asymptotic expressions of the exact results,  the pressure may be written as ($\kappa^2 = 8 \pi \beta n e^2$ \cite{RedBook,GreenBook}
\begin{eqnarray}
\beta p (\beta, n) =  n [1 + \frac{1}{8 \sqrt{2}} n \Lambda_e^3 + O(n^2)]
+ n - \frac{\kappa^3}{24 \pi} [1 - \frac{3 \sqrt{\pi}}{8} (\kappa \lambda)+\frac{3}{10}(\kappa \lambda)^2 + ...]\nonumber\\
- n^2 \gamma [1 - \beta e^2 \kappa] + O(n^{3} \log n)\,.
\label{pressPBL1}
\end{eqnarray}
We introduced here the mean thermal wave length
$$
\lambda = \lambda_{ie} = \frac{\hbar}{2 \mu k_B T}=\frac{\Lambda_e}{2 \sqrt{\pi}}; \qquad \mu = \frac{m_e m_i}{m_e + m_i}; \qquad \gamma = \Lambda_e^3 \sigma_{\rm BPL}(T) = 8 \pi \sqrt{\pi} \lambda_{ie}^3 \sigma_{\rm BPL}(T)
$$
where $\lambda$ is an effective quantum length, being in some correspondence to
the Debye-H\"uckel parameter in the classical theory. We note as a special property of the given temperature regime that he
interaction terms in the pressure depend only on reduced mass $\mu$ of the electron-proton interaction and not
on the other reduced masses.
An equivalent form of eq.(\ref{pressPBL1}) which shows better the general structure is the following
\begin{eqnarray}
\beta p (\beta, n) =  n[1 + \frac{1}{8 \sqrt{2}} n \Lambda^3+ O(n^2)] + n - n \gamma [(1 - \beta e^2 \kappa +  O(\kappa^2)) + ...]\nonumber\\
- \frac{\kappa^3}{24 \pi} [1 - \frac{3}{2} \gamma + O(\gamma^2)][1 - \frac{3 \sqrt{\pi}}{8} (\kappa \lambda)+ \frac{3}{10}(\kappa \lambda)^2 + O(\kappa^3)] + O(n^{3/2})\,.
\label{pressPBL2}
\end{eqnarray}
The expression for the pressure of hydrogen in the limit of small densities, given in the formulae
eqs. (\ref{pressPBL1}) and (\ref{pressPBL2}) has been drawn in Fig. \ref{presse}.
This is a first step to a Saha-type description, since the pressure related to the ideal pressure
is decreasing with the density with a factor proportional to the distribution function,
however we see that our formula fails at higher densities going even to negative values.
Looking at the curves in Fig. \ref{presse} we see that the screening effects and the contributions from the Brillouin-Planck-Larkin function both tend to lower the pressure. This lowering is too large at increasing densities, so we have
to look for contributions limiting this growth. In order to proceed in this direction,
we investigate first the structure of eq. \ref{pressPBL2}, aiming to complete the unfinished series
with respect to $\gamma$ and $\kappa \lambda$.
The structure of eq. (\ref{pressPBL2}) suggests that the expressions in the parenthesis may be
only the first terms of some infinite series with respect to $n \Lambda^3$, $\kappa$ and $\gamma$.
The first series is evidently nothing else than the ideal electron pressure, which may easily be extended to an infinite series using the ideal Fermi pressure.
In order to find the higher orders in the series in $\gamma$ and in $\kappa \lambda$ we will
use a more complicated procedure which is based on fugacity expansions \cite{RedBook,EbPhysica74,Bartsch}
According to our general philosophy, the strong decay
of the pressure observed in second order density expansions (Fig. \ref{presse}) is due to the fact that
important higher order terms corresponding to some infinite series were omitted in the derivation based on the canonical ensemble. Alastuey et al. \cite{AlastueyBaCoMa08,Alastuey}
succeeded in deriving a Saha type expression in summing up an infinite series of terms.
Here we follow a similar idea but go an alternative way through fugacity expansions. As shown much earlier,
fugacity-like expansions are equivalent to mass-action laws. This was used in plasma theory
by Bartsch, Ebeling and others \cite{RedBook,EbPhysica74,EbDubna,Bartsch} and worked out in large detail by
Rogers et al. \cite{Rogers}. We mention that Rogers et al. succeeded to show that extended fugacity expansions
allow an excellent description of measurements for the oscillation modes of the sun \cite{Dappen}.\\
A systematical quantum statistical approach to the pressure including the low as well as the high pressure region
may be given on the basis of Green's function representations of the pressure \cite{RedBook,EKKK76,GreenBook,EbKrKrRo86}.
The Green's function representation works in the grand canonical ensemble which provides us a series
in the fugacities instead of a series in the densities.
Note that $\mu_a^{\rm ex}$ is the excess part of the chemical potential (the part beyond the ideal Boltzmann term)
and that therefore $z_a \rightarrow n_a$ for small densities.
The structure of this series is similar to the density series, we have e.g. similar screening effects,
however there are some differences. e.g. the fugacity series contains more diagrams.
The contributions of bound states of an electron and an ion (atoms)  are contained in the contribution
of order $z_e z_i$, the contribution of molecules is contained in the term of order $z_e^2 z_i^2$.
General expressions for screened fugacity series were written down first by Montroll and Ward,
Vedenov and Larkin and explicit calculations were done by De Witt and Larkin \cite{Vedenov,Larkin,DeWitt}.
Semiclassical fugacity series were given by several workers
\cite{RedBook,Bartsch}.
We are interested here mostly in the bound state effect and may simplify by using approximations allowed for the region $T \le 157 000$ K were the plasma is nondegenerate and where the bound states are dominant. As we have seen
 in the density representation, the differences of the masses of electrons and ions (protons) do not play a role in the region
 $T \ll I \ k_B$ and the plasma behaves asymptotically like a system with equal masses and equal relative De Broglie radii.
In the lowest approximation corresponding to eq. (\ref{pressPBL1}) we find for low temperature H-plasmas the following results \cite{RedBook,GreenBook,wittsaka}
\begin{eqnarray}
\beta p = z_e +  z_i + \frac{\kappa_g^3}{12 \pi} f(\kappa_g \lambda)  + 8 \pi
z_e z_i \lambda^3 \exp[1 + \beta e^2 \kappa_g] \sigma_{\rm BPL}(T) + O(z^{5/2})\,,
\label{PressFug0}
\end{eqnarray}
where the grand-canonical screening quantity is defined by
$\kappa_g^2 = 4 \pi \beta (z_e + z_i) e^2 $ and the quantum-statistical ring
function is a transcendental function
having the series
\begin{equation}
f(x) = \left[1 - \frac{3 \sqrt{\pi}}{16} x  + \frac{1}{10} x^2 + ... \right]\,.
\end{equation}
The densities may be expressed by fugacities
\begin{eqnarray}
n_i = z_i \frac{\partial \beta p}{\partial z_i} = z_i + \frac{\kappa_g^2}{16} \frac{\partial}{\partial \lambda} (\kappa_g \lambda f(\kappa_g \lambda)) +
4 \pi z_e z_i [1 + \beta e^2 \kappa_g + \cdots] \lambda^3 \sigma_{\rm BPL} (T)\cdots\,,
\label{DensFugi}
\end{eqnarray}
\begin{eqnarray}
n_e = z_e \frac{\partial \beta p}{\partial z_e} = z_e + \frac{\kappa_g^2}{16} \frac{\partial}{\partial \lambda} (\kappa_g \lambda f(\kappa_g \lambda)) + 4 \pi z_e z_i  [1 + \beta e^2 \kappa_g + \cdots] \lambda^3 \sigma_{\rm BPL}(T)\cdots\,.
\label{DensFuge}
\end{eqnarray}
We will show by iterations that going from the fugacity variable to the density this result
occurs to be equivalent to eq. (\ref{pressPBL1}). In order to proceed with this complicated system of equations we go to the special case of non-degenerate hydrogen at lower temperatures $T \le I / k_B$ and satisfying
$n \Lambda_e^3 \le 1$. As shown above, in this region the differences
of the electron ion masses does not matter and we my assume $z_e = z_i = z$.
In order to represent the pressure by densities we neglect now in a first step the bound state contributions
assuming $\sigma_{\rm BPL} (T) = 0$. Expressing in this approximation first the fugacities by densities we find
\begin{eqnarray}
\hspace*{-1cm}z = n \exp\left[ - \frac{\beta e^2 \kappa}{2} G(\kappa \lambda)\right]; \qquad
G(x) = \frac{\sqrt{\pi}}{x}\left[1 - \exp(\frac{x^2}{4})\right] \left[1-\Phi(\frac{x}{2})\right]= 1 - \frac{\sqrt{\pi}}{4} x + \frac{1}{6} x^2 + O(x^3)\cdots\,,
\label{FktGx}
\end{eqnarray}
where $\Phi(x)$ is the error function. Then excluding the fugacities from the series step by step we arrive
at the representation
\begin{eqnarray}
\beta p = 2 n - \frac{\kappa^3}{24 \pi} \phi(\kappa \lambda); \qquad \kappa^2 = 8 \pi \beta n e^2\,.
\label{Pressdens0}
\end{eqnarray}
Here $\phi(x)$ is another transcendent function, related to $f(x)$ and $f'(x)$, the so-called pressure ring function which is an analogue of the Debye-H\"uckel ring function \cite{Vedenov,RedBook}.
A full representation of this transcendent function is given by the infinite series
\begin{equation}
\phi(x) = 1 - \frac{\sqrt{\pi}}{3} \sum_{k=1}^{\infty} \frac{(k+1)(k+3) x^{2k-1}}{2^k k!} + \frac{1}{3}\sum_{k=1}^{\infty} \frac{(k+1)(k+3) x^{2k}}{2^k (2k-1)!!}\,.
\end{equation}
In the calculations we used more convenient Pad\'e formulae.
The nonlinear functions $\phi(x)$ and $G(x)$ converge to zero at large arguments what is important for the high-density asymptotics of the pressure. In Fig. \ref{presse} the result of fugacity expansions up to 2nd order virial terms for
$T = 15788$ K and $T =  23682$ K (red and green curves)
are compared with corresponding density expansions. We see that the behaviour of
truncated fugacity expansions seems to be much better than that of truncated density series.
However this is not true in all cases and may lead to serious problems. We may show this on the example of the
ideal Fermi gas. Truncating here the series after quadratic terms the representation of the pressure leads for
$n \Lambda^3 > \sqrt{2}$ to an imaginary pressure.
For this reason we hold the opinion that the a controlled summing up of terms in the
density series may be more reliable. The results we obtained so far may be summarized as follows.
By means of grand-canonical methods we succeeded already to obtain one infinite series in the screening parameter
$\kappa$ instead of the first linear and quadratic terms only we found in the density series,
(see eq.(\ref{pressPBL2})). This summing up a series in $\kappa$ which leads to a saturating function
what improves very much the behavior at larger densities.\\
In our canonical representation eq. (\ref{pressPBL2}) appears also another series expansion in the density parameter $\gamma$. In order sum up this series in $\gamma$, we proceed in a similar way. We study at first the bound state contributions, neglecting the contributions coming from the ring terms and from the
terms due to degeneracy. This yields the simple quadratic equation
\begin{eqnarray}
n = z + 8 \pi z^2 \lambda^3 \sigma_{\rm BPL}(T).
\end{eqnarray}
By the way, this formula shows already that the
fugacities of the electrons and the protons should be rather small in the bound state region where the partition function is large.
The quadratic equation for the fugacities can be easily solved.
This way we find after elimination of the fugacity $z$:
\begin{eqnarray}
z^0 = n A(\gamma); \qquad A(x) = \frac{1}{2x}[\sqrt{1 + 4 x} - 1] ; \qquad \gamma= 8 \pi n \lambda^3 \sigma_{\rm BPL}(T).
\end{eqnarray}
Introducing the zeroth step of iteration $z^0$ into the pressure we find
\begin{eqnarray}
\beta p^0 = n (1 + A(\gamma)).
\end{eqnarray}
This representation includes a nonlinear function $A(x)$ which saturates at large densities
and is closely related to the result of the Saha theory, except that in our version the BPL partition function
appears.
The new representation contains all terms in the density up to $n^2$ as well as several higher order terms in the density contained in the nonlinear Saha function $A(\gamma)$. We will show that this kind of function
allows us to reproduce an ideal Saha-type behavior.
We note the following series expansion and asymptotics of the nonlinear function $A(x)$:
\begin{eqnarray}
A(x) = 1 - x + 2 x^2 -\cdots; \qquad A(x) \simeq \frac{1}{\sqrt{x}} +\cdots\,.
\end{eqnarray}
The saturating behavior of the nonlinear function $A(x)$ in comparison to its linear approximation
has the consequence that in the region of large partition functions the fugacities disappear as
\begin{eqnarray}
z^0 = \frac{1}{\sqrt{8 \pi n \lambda^3 \sigma_{\rm BPL}(T)}}.
\end{eqnarray}
This is important for the understanding, why the fugacity series has in the region of bound states a better convergence as the density series.
The better convergence is due to the fact that the fugacities disappear in the region where
the bound state contribution is large.
The fugacity series contains more terms than the density series, so we may expect that some of the difficulties connected with density expansions, as the strong decrease of the pressure with increasing density shown in Fig. \ref{presse} may be avoided. Indeed,  a representation of the curve
corresponding to eq. (\ref{PressFug0}) shows a more reasonable behavior with increasing density which
is much closer to the Saha-type behavior.
As the curves shown in Fig. \ref{presse} demonstrate, the pressure according to the fugacity expansion goes to saturation. This is due to the fact that the fugacity is not fixed, it decreases with increasing
value of the partition function and this way limits the growth. Evidently the fugacity expansions provide a more correct description of the bound state contributions as demonstrated in Fig. \ref{presse}.
In order to draw some first conclusions: \\
Density as well as fugacity expansions have both advantages as well as disadvantages:
1) The density expansion describes well the screening effects but
it fails to cope with the diverging contributions from the screening terms and from the BPL-partition function.
2) The fugacity expansion corresponds to an infinite density series including the partition function $\sigma$.
If $\sigma$ is large, then the fugacity goes to zero what guarantees always
finite contributions to the pressure. This is also true for the screening contributions; any strong increase of contributions damps the fugacities.\\
We may expect that the best representation is obtained by extended density expansions
which contain additional contributions corresponding to the important damping terms in the
fugacity expansions and provide Saha-like terms.
This kind of mixed expansions combine the positive features of both expansions
avoiding the negative features. In order to demonstrate this we started here from the density expansions and used
the fugacity expansions for finding the right continuation of the infinite
series with respect to the $\gamma$- parameter and the $\kappa \lambda$ parameter. \\
This way we obtain in a first step the following expression for the fugacity and the pressure
\begin{eqnarray}
z = n A \left(\exp[ - \beta e^2 \kappa G(\kappa \lambda) ] \gamma \right)\,,
\label{Fugdens1}
\end{eqnarray}
\begin{eqnarray}
\beta p =  n -  \frac{\kappa'^{3/2}}{24 \pi} \phi(\kappa' \lambda) +
n A \left( \exp[ - \beta e^2 \kappa G(\kappa' \lambda)] \gamma \right), \qquad \kappa' = \kappa A(\gamma)^{1/2},
\label{presscom}
\end{eqnarray}
where a renormalized value of the screening parameter $\kappa'$ appears.
This provides us a closed and relatively simple formula for the pressure
which however is restricted to the nondegenerate region.
However so far our results refer to non-degenerate electrons.
The good convergence of the new expression including nonlinear functions $A(x)$ and $\phi(x)$ based on the
fugacity series is explained by the fact that the fugacities of the electrons and the protons are rather small in the
bound state region. The overall behavior of
the new representation is much better than that of pure density or pure fugacity representations.
We may conclude that the most appropriate description of Coulombic systems is by density expansions and including
infinite sums representing screening and bound state effects.
We note that the new theory based on eq. (\ref{presscom}) is consistent with the Saha-theory and in particular also with the
 Saha-Debye-H\"uckel theory \cite{RedBook}.
We underline that several of the terms beyond $n^2$ and $z^2$ are based on extrapolations which
still need further confirmations. However in the region of low temperature $T < I / k_B$ and
non-degenerate plasmas our rather simple formulae give a rather good behavior
and describe well the transition from low density to the valley of bound states.
We note however that several physical effects as e.g. plasma phase transitions \cite{RedBook,EbRi85}
are not yet described by the present approach. Maybe this effect appears only after
transition to some chemical picture \cite{RedBook,EbRi85}.\\
In order to describe as well the transition to full ionization in the degenerate region,
additional effects are to be taken into account, in particular the symmetry between
electrons and protons is lost. The most important new effects at high densities are connected of the replacement
of the ring terms by Hartree-Fock terms and a density dependence and final disappearance of the
bound states with increasing density.

\section{Calculation of the pressure at high densities in mean field approximation}

\begin{figure}[htb]
\begin{center}
\includegraphics[width=4cm,angle=-90]{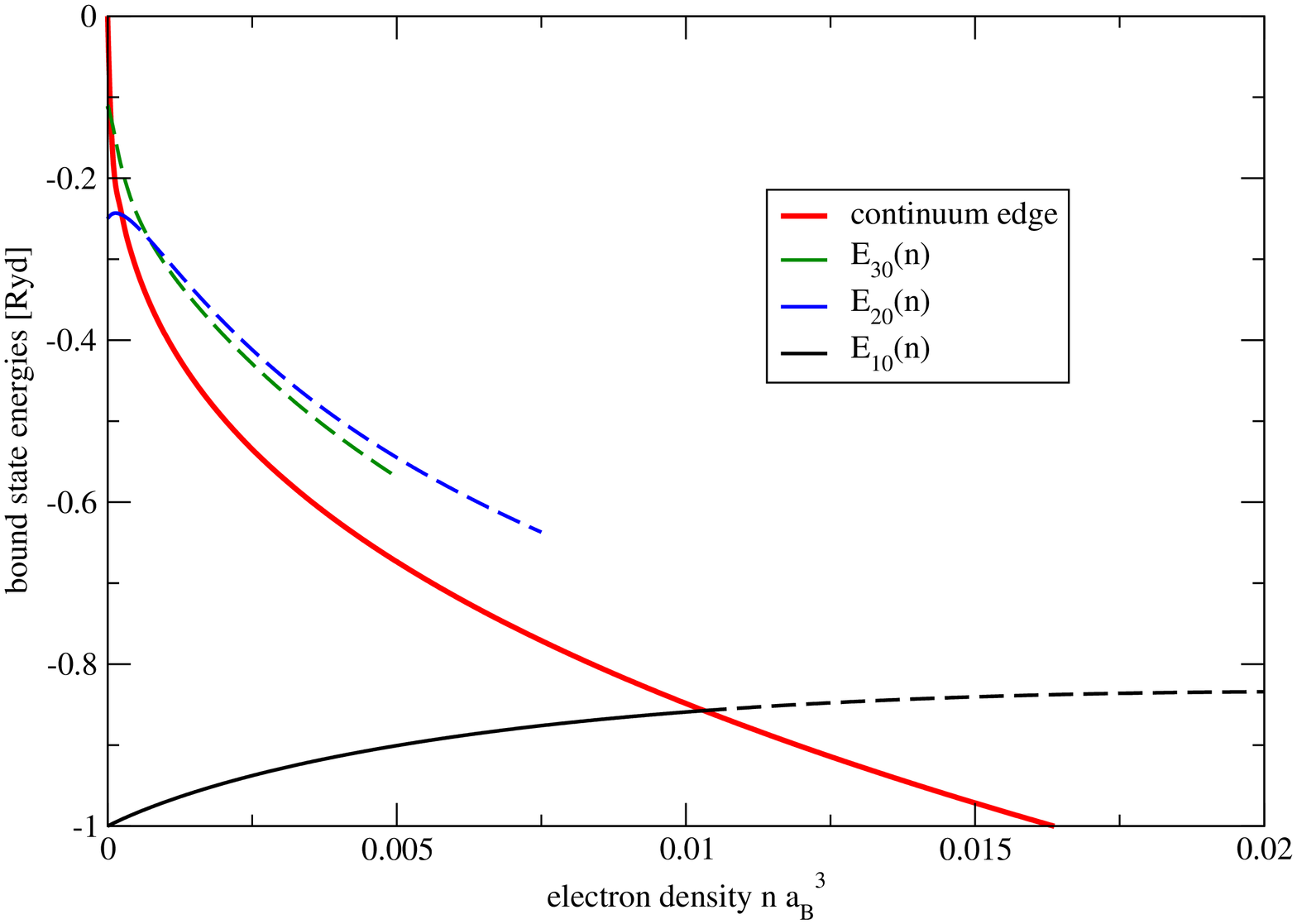}
\includegraphics[width=4cm,angle=-90]{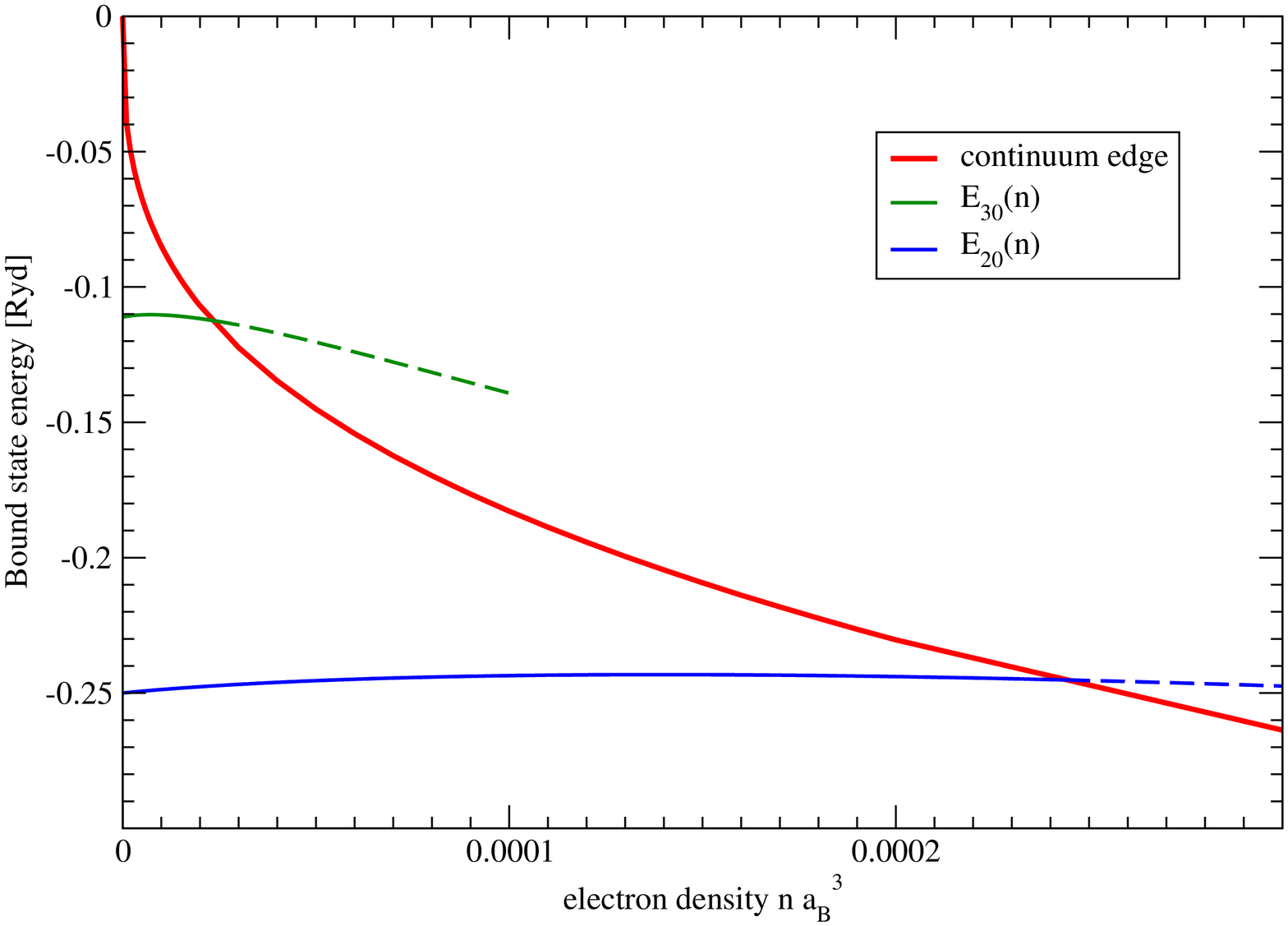}
\end{center}
\caption{The shifts of the ground state and two excited states
in Pauli-Fock approximation in relation to the lowering of the continuum
 level.}
 \label{shifts}
\end{figure}

The theory presented in the previous section provides isotherms of the pressure for the nondegenerate
region. After crossing the line of degeneracy $n \Lambda_e^3 \simeq 1$ the theory is no longer valid since degeneracy effects and the dissymmetry of the massed are relevant. In this region the formulae given above are just an extrapolation which is exact only in the (dominant) ideal Fermi contributions. For very dense, strongly degenerate plasmas, bound states do not exist, the plasma behaves like a degenerate nonideal Coulomb gas.
Let us assume, we start from a high density and decrease the density along an isotherm.
The question is then, at which density bound states will appear. In other words we have to ask the question
at which densities the spectrum will have discrete levels corresponding to bound states of two particles, three particles, four particles etc.
Turning the question around we may ask, what is the density dependence of discrete states.
This way, the most important effect to be taken into account in the degenerate region
$n \Lambda_e^3 > 1$ is the change of the energy levels by effective density-dependent energy levels
and the question of merging of discrete levels in the continuum. This strongly affects the equation of state of hydrogen \cite{GreenBook,RKKK78}.\\
In the region of strongly degenerate electrons, the thermodynamics is in good approximation described by the
Hartree-Fock approximation, the protons form a kind of a Wigner lattice \cite{GreenBook}. Based on eq. (\ref{qupress}) the grand-canonical pressure may be approximated by \cite{RedBook,GreenBook,EbKrKrRo86}
\begin{eqnarray}
\label{hdpress}
p(\beta, z_e, z_i)& =&
p^{\rm id} + p_{e}^{\rm HF}(z_e) + p_{i}^{\rm Wi}(z_i) + z_e z_i \Lambda_e^3 {\tilde \sigma}(T) + ...\\
{\tilde \sigma}(T)& = &\sum \left[\exp(-\beta {\tilde E}_n) -1 + \beta {\tilde E}_n \right]
\end{eqnarray}
with
\begin{eqnarray}
\label{hdfug}
n_e = z_e + z_e \frac{\partial \beta p_{e}^{\rm HF}(z_e)}{\partial z_e} + z_e z_i \Lambda_e^3 {\tilde \sigma}(T)\,,\\
n_i = z_i + z_i \frac{\partial \beta p_{i}^{\rm Wi}(z_i)}{\partial z_i} + z_e z_i \Lambda_e^3 {\tilde \sigma}(T)\,.
\end{eqnarray}
We proceed as in the previous section. Neglecting first the bound state contribution we get for the
fugacity of the electrons in Hartree-Fock approximation \cite{RedBook,GreenBook}
\begin{eqnarray}
\label{elfug}
z_e = n \exp[- \beta \frac{e^2}{\Lambda_e} I_{-1/2}(\alpha)]; \qquad \alpha = \beta \mu_e^{\rm id}.
\end{eqnarray}
Here the $I_k(x)$ are the well-known Fermi integrals. We use in the following the high density limit
including the first temperature corrections calculated with the Sommerfeld method
\cite{RedBook,EKKK76,GreenBook,KrSt79} ($t = k_B T / {\rm Ry}$).
Since the $T^2$ correction in the Sommerfeld expansion changes the sign with increasing temperature we used in the calculations
a more convenient Pad\'e expression with the $T^2-$ term in the denominator
\begin{eqnarray}
\label{SoFeld2}
p_{e}^{\rm HF} [{\rm Ry}] \simeq -\frac{0.3059 \,n}{r_s + 6.030\, t^2 \,r_s^5}; \qquad z_e^{\rm HF} \simeq
n \exp[-\frac{1.222}{t \,r_s + 0.1216\, t^3 \,r_s^5}]\,, \qquad r_s =\left[\frac{3}{4\pi n}\right]^{1/3}\,.
\end{eqnarray}
The ionic contribution at high density is more difficult \cite{GreenBook}. However in the limit of high density,
the Wigner limit, we may give the result in an analytically quite similar form with lowest order $T$-corrections:
\begin{eqnarray}
\label{iopresshd}
z_i \simeq n \exp[-\frac{2.3856\, (1 + 0.05 \,r_s)}{t \,r_s}] ; \qquad  p_{i}^{\rm Wi} [{\rm Ry}] \simeq \frac{0.5964\, n (1 + 0.01\, r_s)}{r_s}\,.
\end{eqnarray}
In difference to the situation at small densities, the fugacities are at high densities (small $r_s$)
rather small in comparison to the densities. For this reason the contribution of the bound states to the pressure
is just a perturbation and we do not need to sum up an infinite Saha-type series, as in the low density case.
Putting now together all terms we arrive in the canonical ensemble at the following simple relation for the
high density pressure of hydrogen ($p_{e}^{\rm F}$- Fermi pressure of the electrons)
\begin{eqnarray}
p(\beta, n) \simeq
p_{e}^{\rm F} + p_{e}^{\rm HF} + p_{i}^{\rm Wi} + z_{e}^{\rm HF} z_{i}^{\rm Wi} \Lambda_e^2
\sum \left[\exp(-\beta {\tilde E}_n) -1 + \beta {\tilde E}_n \right]\,.
\end{eqnarray}
This way we found now for the limit of high density a closed expression for the pressure of hydrogen in the physical picture. It remains to
find the effective two-particle energy levels which are now density-dependent.
The two-particle bound state energies in a dense system $\tilde{E_{n}}$ may be derived from an effective
wave equation, the so-called Bethe-Salpeter equation.\\
In this paper, the influence of the medium is taken only in mean-field approximation, i.e., by Pauli blocking and Fock self energy . The
effective wave equation reads in this case (in the following the tilde is omitted)
\cite{GreenBook,RKKK78,Zimmetal}
\begin{equation}
\label{inmedium}
\frac{p^2}{2m_e} \psi_n(p)+ \sum_q V(q) \psi_n(p+q) - E_{n} \psi_n(p) =
\sum_q V(q) \left[ \psi_n(p+q) f_e(p) - \psi_n(p) f_e(p+q) \right]\,.
\end{equation}
Here the l.h.s. represents the standard Schr\"odinger equation
in momentum representation. The terms on the r.h.s. containing the electron distribution function
$f_e(p)$ model effects connected with the Fermi character of the electrons.
The first of these two contributions we denote as Pauli-blocking and the second one as
Fock contribution. The Fock term
and the Pauli blocking contributions have opposite signs and compensate each other partially.
The more complete Bethe-Salpeter equation contains also
dynamic self-energy and dynamically screened interaction contributions. These corrections were studied for hydrogen and electron-hole systems, see, e.g., \cite{GreenBook,RKKK78,Zimmetal,Arndt96,KKKWitt,Fehr,Seidel,Bornath99}.\\
The corresponding shifts
will be neglected here since we consider only the high density limit.
In perturbation theory, the sum of Pauli and Fock shifts \cite{KSKBook,Fehr,EbBlRRR09} can be calculated with the solutions $\phi_n$ of the unperturbed Schr\"odinger equation.
\begin{equation}
\Delta E_{n}^{\rm PF} = - \sum_{p,q} \phi_n(p) V(q) [f_e(p) \phi_n(p+q)- f_e(p+q) \phi_n(p)].
\label{paulifocknonlin}
\end{equation}
In the limit of strong degeneracy where the Fermi function can be replaced by the Heaviside step function,
the integrals can be performed analytically. The result for the shift of the lowest three energy eigenvalues is shown in Fig. \ref{shifts}.
In the low-density limit we have for the ground state $n=\{10\}$
\begin{equation}
\lim_{n \to 0} \Delta E_{10}^{\rm PF}= \frac{n}{2} \sum_q \frac{4 \pi e^2}{q^2} \phi_{10}(q) [\phi_{10}(0) - \phi_{10}(q)] =
32 \pi n' -20 \pi n' =12 \pi n' ~.
\label{paulifock}
\end{equation}
In analogy we find analytical expressions for the excited $s$-states. In the low-density limit we have
\begin{equation}
\lim_{n \to 0} \Delta E_{20}^{\rm PF} = 48 \pi n'\,,\qquad \lim_{n \to 0} \Delta E_{30}^{\rm PF} = 108 \pi n'\,.
\end{equation}
As we see, due to compensation effects,
the sum of Pauli and Fock shifts is, in linear approximation, less than a half
of the Pauli shift in the ground state and is also reduced in the excited
states. The evaluation of eq. (\ref{paulifocknonlin}) for arbitrary densities leads to deviations from the linear density dependence as shown in Fig. \ref{shifts}. In contrast to the bound states, there is no Pauli shift for the scattering states. The lowering of the continuum edge is  given
by the Fock energy shift at zero momentum,
\begin{equation}
\Delta E^{\rm Fock}(p=0) = - \sum_q V(q) f_e (q) = - \frac{4 p_F}{\pi} = - 4 \left(\frac{3 n'}{\pi}\right)^{1/3}.
\end{equation}
The Fock shift at the Fermi momentum $\Delta E^{\rm Fock}(p_F)$ results as $\Delta E^{\rm Fock}(p=0)/2$.  This is in agreement with the estimate of the shift of the continuum by the chemical potentials
given above since in the strongly degenerate case the chemical potential of the electrons is given by the energy at the Fermi momentum. The lowering of the continuum edge is also shown in Fig. \ref{shifts}. The
bound states vanish when they meet the continuum edge.
We see that for $n' < 0.0018$, only two levels survive and
for
$n' < 0.00018$ still 3 levels are left. Only at very small densities more than 3 levels
are left.
We repeat here that the solution of an in-medium Schr\"odinger more complex than Eq.(\ref{inmedium})
was dealt with in \cite{GreenBook,RKKK78,Zimmetal,Arndt96,KKKWitt,Fehr,Seidel,Bornath99}.
In \cite{Fehr} it was shown that perturbation procedures are justified  even for in-medium Schr\"odinger equations.  Solutions beyond perturbation theory can also be found within a variational approach \cite{MottBook,EbBlRRR09}.

\begin{figure}[htb]
\begin{center}
\includegraphics[width=5cm,angle=0]{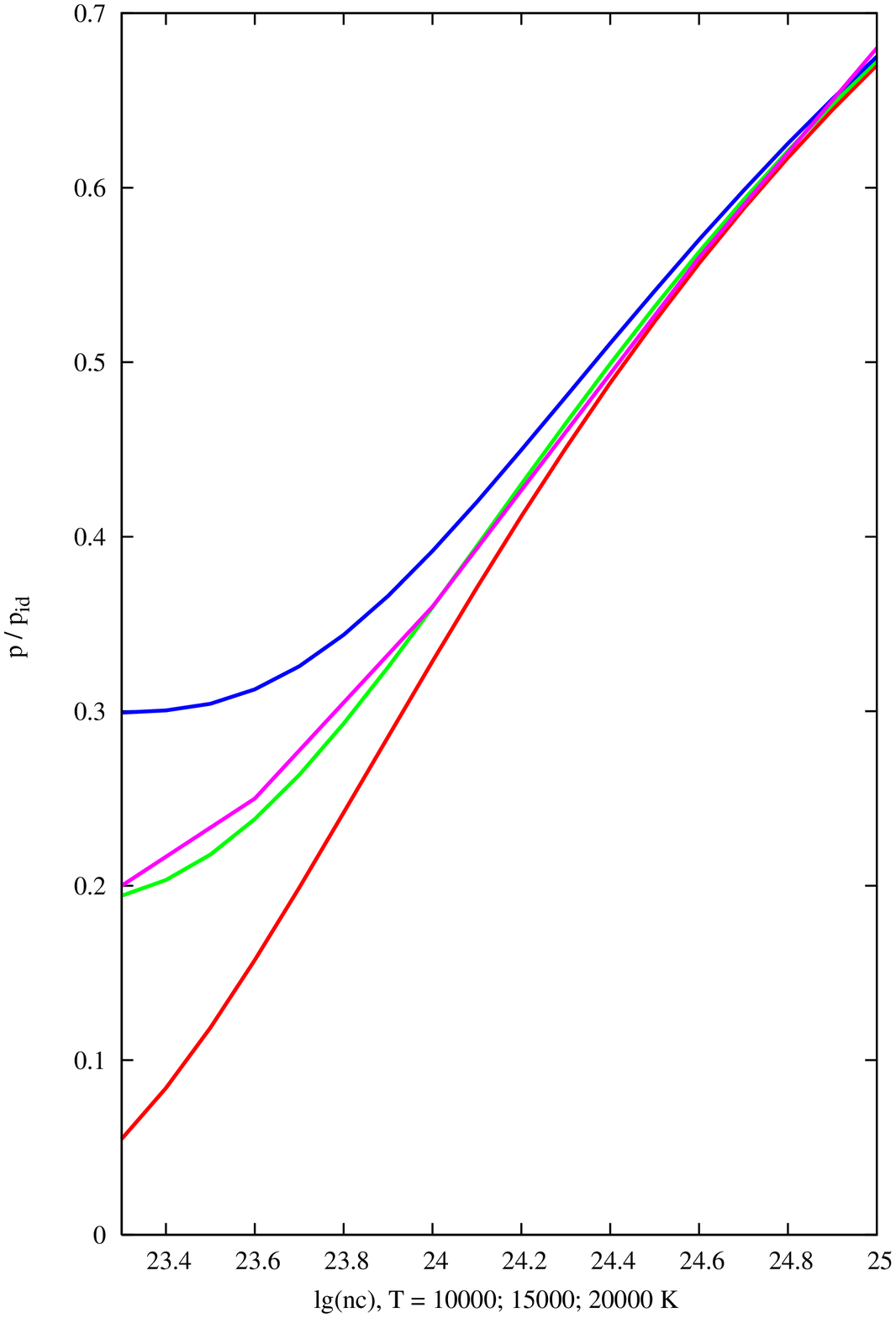}
\includegraphics[width=5cm,angle=0]{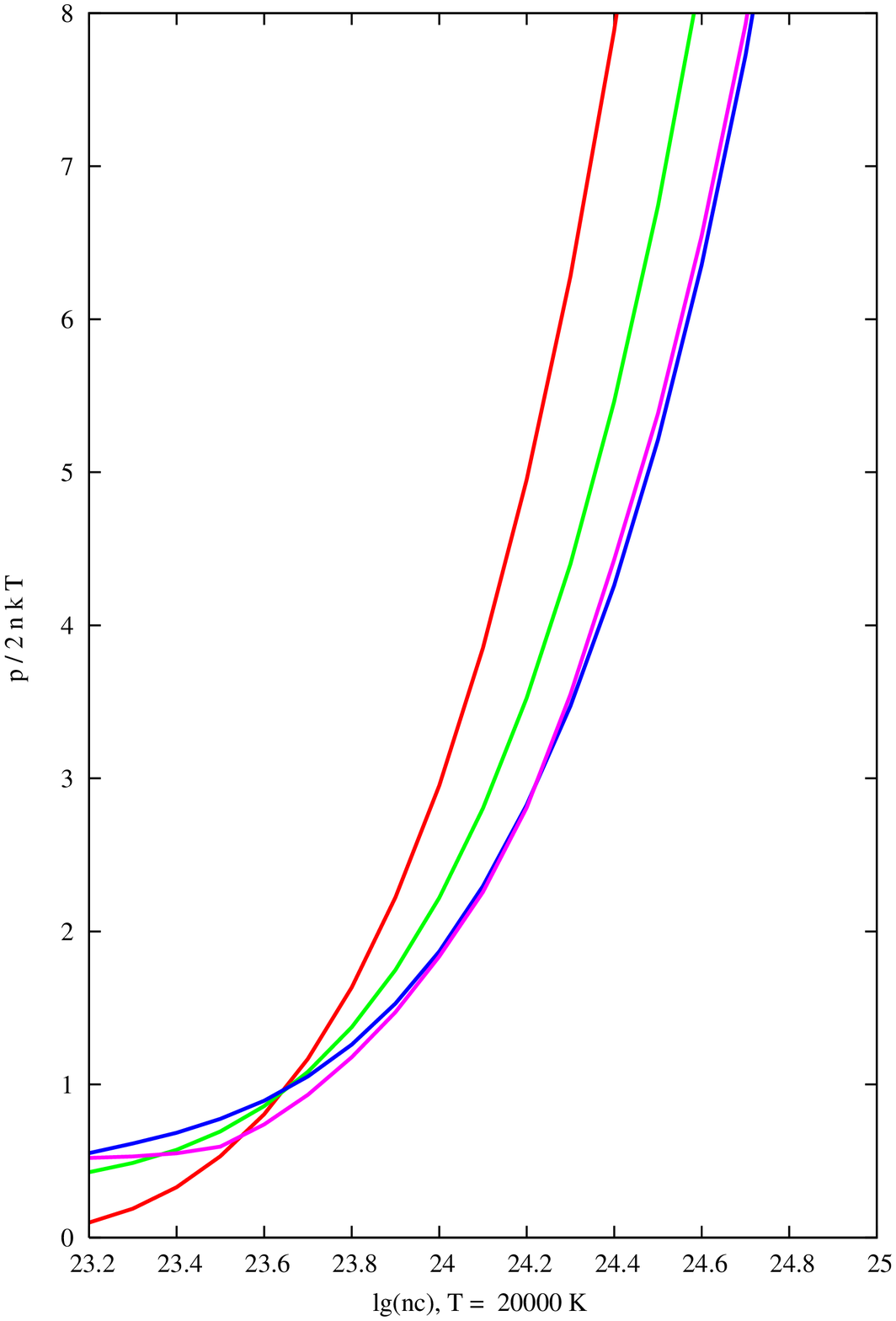}
\end{center}
\caption{ Calculations of the pressure corresponding to the Hartree-Fock approximation taking into account the energy shifts for the case of high densities.
The left panel shows $p / p_{id}$ for the temperatures
$T = 10000, 15000, 20000$ K. For ${\rm log n} < 23.5$ begins the region of developed bound states which is not well described
by this method. The magenta curve represents numerical and analytical results by Vorberger et al. \cite{Gericke11}
for $T = 10000$ K.
In the right panel we show the pressure related to the classical ideal pressure at the same temperatures
and for comparison we give for $T = 20000$ K a curve (in magenta) obtained earlier using a Pad\'e approximation
for the pressure in combination with a nonideal Saha equation \cite{Pade90}.
}
\label{PHighdens}
\end{figure}

As an application we calculated the pressure related to the classical ideal
pressure for the case of high densities at a temperature of $10 000 - 20 000$ K,
corresponding to the Hartree-Fock and Wigner approximations, see Fig. \ref{PHighdens}. The energy shifts
are taken into account. For comparison we have given a curve for 10 000 K given
by Vorberger et al. \cite{Gericke11} and with a result for $T = 20 000$ K
obtained by means of a Pad\'e approximation in combination with a mass action law
\cite{Pade90}. The agreement of our new, rather simple approach with the
previous ones is rather satisfactory.

\section{Discussion and Conclusion}

We discuss here the divergence of the partition function of the Bohr atom
and the BPL method of renormalization, and present several new results within the physical picture.
At small densities we discussed closed expressions for the ring sums. At high densities we
concentrated on the Hartree-Fock approximation and took
into account the shifts in Hartree-Fock approximation. We assume that these shifts
provide the most important effects for the destruction of bound states in the high-density region.
In particular we contribute here to the theory of hydrogen at high pressures in the region
where a Mott transition to full ionization  has been predicted and where
recent experiments have shown a transition from insulating behavior to
metal-like conductivity.

In order to understand this transition
several effects have to be taken into account. We concentrated here on
so-called Pauli blocking effects expressing the rule that states occupied by
atomic electrons cannot be occupied by free electrons with the same spin state.
This leads at high electron densities to the destruction of
atomic states which need a relatively high amount of phase space.
We calculated the energy shifts due to Pauli effects and discuss
the Mott effects solving effective Schr\"odinger equations
for strongly correlated systems.

Still a word on the plasma phase transition. In the temperature region $10000 - 20000$ K which we studied here,
no first order transitions were observed. The physical picture which we used here cannot be extended without problems to lower temperatures, where ${\rm H}_2$ molecules dominate. In order to check for phase transitions we introduced the energy levels calculated here into the
chemical picture developed earlier \cite{MottBook} which includes ${\rm H}_2$ molecules.
A resulting isotherm for $T = 6000$ K shows a small wiggle. According to this estimate the transition
occurs in the range below 6000 K, the critical temperature $T_c$ is slightly above.
Summarizing, we show here that in the non-degenerate the most essential effects are the screening and the
BPL reduction, resulting in absorbing the orders in $e^2,\,e^4$ and the compensation effects of the states below and above the series limit. Another important effect is the summation of an infinite series in the
parameter $\gamma = \Lambda^3 \sigma_{\rm BPL}(T)$ leading to a Saha-type behavior.
At high density the Pauli-Fock contributions are most essential for the limitation of the number of energy levels
and determine the transition from partial ionization  to full ionization when densities become very high .

\end{document}